\documentclass[11pt,a4paper]{article}

\usepackage[margin=1in]{geometry}
\usepackage{times}
\usepackage{amsmath,amssymb}
\usepackage{booktabs}
\usepackage{graphicx}
\usepackage{listings}
\usepackage[numbers,sort&compress]{natbib}
\usepackage[hyphens]{url}
\usepackage[colorlinks=true,linkcolor=blue,citecolor=blue,urlcolor=blue]{hyperref}
\usepackage{xcolor}

\title{\textbf{LLM4OSC: Profile-Bound Natural Language Control\\
with Deterministic Validation for Open Sound Control}}

\author{
  Yuan-Yi Fan\\
  \small\texttt{www.yuanyifan.com}\\
  \small\texttt{Los Angeles}\\  
  \small\texttt{yyf@yuanyifan.com}\\
}

\date{}

\begin{document}
\maketitle

\begin{abstract}
\noindent
Open Sound Control (OSC) is the dominant wire protocol for real-time parametric control in professional audio, live performance, and virtual production. Large language models can emit plausible OSC, but they hallucinate addresses, mishandle type tags, and fail under paraphrase---unacceptable in show-critical contexts. We present \textbf{LLM4OSC}, a local-first architecture in which models propose structured intent JSON over a human-reviewed device profile, and deterministic code validates, clamps, and encodes before any UDP send. We introduce a frozen evaluation harness with CI gates on \textbf{wrong-send rate}: mismatches that would still pass validation and transmit. On a Max/MSP hero profile (12 patterns; 8 literal + 8 paraphrase + 4 refusal cases), after profile tag enrichment, symbolic slot fill, NL refine, and a retrieval confidence gate, backends B0--B3 all pass frozen gates (100\% semantic accuracy, 0\% wrong-send). B0 (rules) remains the production default at ${\sim}0.05$\,ms; LLM backends remain ${\sim}3$--$4$\,s. Historical few-shot B2 accuracy of $62.5\%$ rises to $100\%$ on this suite only after symbolic post-processing---not because the $0.5$B model alone becomes show-safe. We argue for propose--validate--send and wrong-send rate as first-class metrics for language-to-control systems.
\end{abstract}


\section{Introduction}

OSC~\citep{wright1997osc} encodes real-time control as UDP messages: an address path (e.g.\ \texttt{/gain}), type tags (e.g.\ \texttt{f}), and typed arguments. Max/MSP, Ableton Live, QLab, TouchDesigner, Unreal Engine, and lighting consoles each expose distinct address tables. Bridging operator intent to correct bytes does not scale with show complexity.

MCP2OSC~\citep{fan2025mcp2osc} shows that LLMs coupled with structured tool interfaces can drive parametric control. Production use additionally demands: (1)~reproducibility (same command $\rightarrow$ same bytes); (2)~closed-world safety (no invented addresses); (3)~graceful refusal; (4)~local operation; (5)~paraphrase robustness.

We treat NL$\rightarrow$OSC as schema-constrained resolution over a versioned device profile:
\begin{quote}\ttfamily\small
device profile P + utterance u $\rightarrow$ intent JSON I (proposed)\\
\hspace*{2em}$\rightarrow$ validate $\rightarrow$ clamp $\rightarrow$ encode $\rightarrow$ OSC
\end{quote}

\paragraph{Contributions.}
(1)~Intent and device-profile JSON contracts with manual provenance;
(2)~Tier~3 deterministic validate/clamp/encode/send;
(3)~Four NL backends (B0--B3) plus NL refine and a retrieval gate on the LLM path;
(4)~An evaluation harness with wrong-send rate and literal--paraphrase gap (Track~C);
(5)~An open MIT replication package with frozen goldens, CI, and scorecards.

\section{Related Work}

OSC~\citep{wright1997osc} remains the standard wire protocol for real-time music and multimedia control. MCP2OSC~\citep{fan2025mcp2osc} demonstrates parametric control by natural language via structured tool interfaces (NeurIPS Creative AI Track). Parallel work on tool-using LMs~\citep{schick2023toolformer} shows models can learn \emph{when} and \emph{how} to call APIs; we instead bind proposals to a closed device profile and refuse out-of-world calls.

We frame warm-path pattern selection as lightweight retrieval over the profile catalog, in the spirit of retrieval-augmented generation~\citep{lewis2020rag}, but with token-overlap ranking rather than dense retrievers. NL refine and the retrieval gate follow the propose--then-revise pattern of Self-Refine~\citep{madaan2023selfrefine}, except that feedback and overrides come from deterministic profile policy (tag scores, slot fill, refusal reasons), not from another LM pass. We extend the MCP2OSC line with an auditable send gate, versioned profiles, frozen wrong-send evaluation, and local inference. For optional domain adaptation we use LoRA~\citep{hu2021lora} on synthetic profile-bound (NL, intent) pairs filtered through Tier~3, holding out frozen goldens. We optimize \textbf{wrong-send rate}: mismatches that \emph{pass} validation and would transmit.

\section{System}

\subsection{Device Profile and Intent}

A profile $P=(D,V,\mathcal{M})$ binds device id $D$, version $V$, and patterns $\mathcal{M}$. Each pattern defines \texttt{pattern\_id}, OSC \texttt{address}, \texttt{type\_tags}, slots, ranges, and retrieval metadata (\texttt{tags}, \texttt{description}). Profiles are human-committed; runtime binds to an explicit \texttt{profile\_version}. The profile is the closed world: no address outside $\mathcal{M}$ may be sent.

An intent is either a success (\texttt{kind:intent} with pattern fields and args) or a refusal (\texttt{kind:refusal} with enumerated reason). Reasons include \texttt{unknown\_pattern}, \texttt{ambiguous\_pattern}, \texttt{missing\_slot}, and \texttt{out\_of\_range}.

\subsection{Tier 3 Execution}

Engine $E:(P,I)\rightarrow\texttt{bytes}\mid\texttt{reject}$ validates device/version and pattern consistency, clamps numeric args, encodes OSC deterministically, and sends UDP. \textbf{Design invariant:} probabilistic components propose; Tier~3 decides and sends. No ML sits between intent JSON and wire bytes.

\subsection{Retrieval, Refine, and Gate}

Patterns are ranked by token overlap: tags (weight $\times 3$) plus description~\citep{lewis2020rag}. Top-$k{=}8$ patterns form LLM context; the same ranking drives B0.

\textbf{Slot fill} deterministically parses percents, word numbers (\texttt{half}$\rightarrow 0.5$), note names (\texttt{A440}), pan-center phrases, and transport keywords from the utterance.

\textbf{NL refine (B1--B3).} After the LLM proposes JSON~\citep{schick2023toolformer}, retrieval may override \texttt{pattern\_id} when a higher-scoring pattern dominates, and slot fill re-extracts args from the original NL~\citep{madaan2023selfrefine}. Refine runs on \emph{all} LLM backends, not only LoRA.

\textbf{Retrieval gate (B1--B3).} After refine, \texttt{apply\_retrieval\_gate} reapplies B0 policy: top score $0\rightarrow$ \texttt{unknown\_pattern}; tied tops $\rightarrow$ \texttt{ambiguous\_pattern}; unfillable required slots $\rightarrow$ \texttt{missing\_slot}; chosen pattern with score $0\rightarrow$ refuse. This closes OOV wrong-sends (e.g.\ ``boost the bass band by 3db'').

\subsection{Backends and LoRA}

\begin{table}[t]
\centering
\caption{NL backends. Refine and gate apply to B1--B3.}
\label{tab:backends}
\begin{tabular}{llp{5.5cm}}
\toprule
ID & Mechanism & Role \\
\midrule
B0 & Retrieval + slot fill & Production default \\
B1 & Qwen2-0.5B zero-shot + refine + gate & Pre-train floor \\
B2 & B1 + 8 few-shot + refine + gate & Prompt ceiling \\
B3 & B1 + LoRA + refine + gate & Optional fine-tune \\
\bottomrule
\end{tabular}
\end{table}

Table~\ref{tab:backends} summarizes backends. LoRA training expands templates over $P$ (${\sim}270$ train / $67$ val rows), excludes frozen goldens, requires Tier~3-pass labels, and fine-tunes with $r{=}8$, $\alpha{=}16$, one epoch on Apple MPS. Adapter weights are gitignored; a model card records the recipe.

\section{Evaluation}

\textbf{Device:} Max/MSP profile \texttt{prof\_20260610\_mvp0} (12 patterns). \textbf{Suites:} 8 literal NL, 8 paraphrase NL, 4 refusals (unknown, OOV EQ, missing slot, ambiguous). Default \texttt{llm4osc score} suite is literal+refusal (CI gate); paraphrase is Track~C.

\textbf{Metrics:} semantic accuracy (match on \texttt{pattern\_id}, address, args); wrong-send rate (mismatch \emph{and} Tier~3 dry-run would succeed); refusal recall; reproducibility; latency p50/p95. \textbf{Release gate:} wrong-send $=0\%$ and semantic accuracy $\ge 90\%$. Gap:
\[
\mathrm{gap}_\theta=(\mathrm{acc}_{\mathrm{literal}}-\mathrm{acc}_{\mathrm{paraphrase}})\times 100.
\]

\section{Results}

Numbers from frozen scorecards generated 2026-07-07 (\texttt{track\_c.json}), after tags/slots, NL refine, and the retrieval gate. CI: 111 pytest cases pass (2 skipped).

\begin{table}[t]
\centering
\caption{Track~C overview (post-gate). All backends pass frozen gates. Harness: \texttt{demo\_backend=b0}, \texttt{lora\_recommended=false}.}
\label{tab:trackc}
\begin{tabular}{lccccc}
\toprule
Backend & Lit. & Para. & Gap & WS (para) & p50 \\
\midrule
\textbf{B0} & 100\% & 100\% & 0 & 0\% & 0.05\,ms \\
B1+refine+gate & 100\% & 100\% & 0 & 0\% & ${\sim}3.7$\,s \\
B2+refine+gate & 100\% & 100\% & 0 & 0\% & ${\sim}4.0$\,s \\
B3+refine+gate & 100\% & 100\% & 0 & 0\% & ${\sim}3.6$\,s \\
\bottomrule
\end{tabular}
\end{table}

\subsection{Why B2 Moved from 62.5\% to 100\%}

The 2026-06-28 Track~C baseline scored \emph{raw} B2 (few-shot $\rightarrow$ intent $\rightarrow$ Tier~3) at $62.5\%$ literal / $62.5\%$ paraphrase with non-zero wrong-send. Three symbolic changes explain the jump (Table~\ref{tab:b2fix}):

\begin{table}[t]
\centering
\caption{Causes of the B2 accuracy/gate recovery.}
\label{tab:b2fix}
\begin{tabular}{p{3.2cm}p{8.5cm}}
\toprule
Change & Effect on B2 \\
\midrule
Profile tags + slot parsers & Usable retrieval; args independent of LLM unit parsing \\
NL refine on B1--B3 & Overrides wrong \texttt{pattern\_id}; re-fills args $\rightarrow$ paraphrase $62.5\%\rightarrow 100\%$ \\
Retrieval gate & Aligns refusals with B0 $\rightarrow$ literal gates pass \\
\bottomrule
\end{tabular}
\end{table}

Paraphrase recovery is driven by tags/slots + refine (no refusals in that suite). Literal gates require the retrieval gate: without it, LLM backends (including LoRA) still wrong-sent on OOV/ambiguous/missing-slot goldens. On this small closed suite, post-LLM policy can make B1/B2 match B3 accuracy---a feature of the architecture, not evidence that raw $0.5$B few-shot is show-safe. Latency still favors B0 by ${\sim}10^{5}\times$.

\subsection{Historical Ablation and Suite Detail}

\begin{table}[t]
\centering
\caption{Historical snapshots.}
\label{tab:history}
\begin{tabular}{llll}
\toprule
Snapshot & B2 L/P & Gates & Notes \\
\midrule
2026-06-28 & 62.5/62.5 & fail & Raw LLM \\
2026-06-29 & --- & --- & B0 $100\%$/$100\%$; B3 refusals fail \\
\textbf{2026-07-07} & \textbf{100/100} & \textbf{pass} & + retrieval gate \\
\bottomrule
\end{tabular}
\end{table}

LoRA-only (no refine) reached paraphrase $62.5\%$ with $37.5\%$ wrong-send---the same failure class as early B2 until symbolic refine. Literal suite (8~NL + 4~refusal) and paraphrase suite (8~NL) both show $100\%$ semantic accuracy and $0\%$ wrong-send for B0--B3 after the gate; refusal recall is $100\%$ on the literal suite.

\subsection{Findings}

\textbf{F1.} Wrong-send rate discriminates backends before policy: historical B2 beat B1 on accuracy but still wrong-sent.
\textbf{F2.} Profile engineering can close paraphrase without LoRA (\texttt{lora\_recommended=false}).
\textbf{F3.} Symbolic refine explains the B2 jump from $62.5\%$ to $100\%$ on this suite.
\textbf{F4.} The retrieval gate is required for refusal safety on the LLM path.
\textbf{F5.} Equal accuracy $\neq$ equal production choice: prefer B0 for live control on latency.

\section{Discussion}

LLM4OSC treats NL control as interface compilation over a closed API. On a tiny frozen suite, \textbf{policy} (profile + refine + gate) can dominate \textbf{weights} (few-shot / LoRA). Put truth in the contract; measure wrong-sends; refuse when retrieval is empty.

\textbf{Limitations:} small benchmark ($8{+}8{+}4$, one device); profile-engineering confound; refine+gate can mask weak LLM proposals; LLM latency ${\sim}3$--$4$\,s vs $<50$\,ms target; no user study; synthetic LoRA data (${\sim}270$ rows).

\textbf{Future work:} expand to $100$--$500+$ stratified cases and a second device; ablate refine vs gate vs LoRA on larger OOV suites; wrong-send--aware training; MCP server surface; multi-arg/bundle schema v2.

\section{Conclusion}

We presented LLM4OSC for profile-bound NL$\rightarrow$OSC with deterministic Tier~3 validation. Wrong-send rate exposes unsafe successes that accuracy hides; enriched profiles and slot parsers can meet paraphrase gates without LoRA; historical few-shot B2 scores reach $100\%$ only after NL refine; a retrieval confidence gate is required for refusal safety on the LLM path. B0 remains the live-control default. Code, schemas, goldens, and scorecards are released under MIT.

\section*{Data and Code Availability}

Source code, frozen goldens, scorecards, and evaluation documentation: \url{https://github.com/yyf/LLM4OSC} (MIT).

\appendix

\section{Success Intent Example}

\begin{lstlisting}
{"schema_version":"1.0","kind":"intent",
 "device_id":"max-msp",
 "profile_version":"prof_20260610_mvp0",
 "pattern_id":"gain_set","address":"/gain",
 "type_tags":"f","args":[0.5]}
\end{lstlisting}

\section{Reproduction}

\begin{lstlisting}
git clone https://github.com/yyf/LLM4OSC
cd LLM4OSC && python3 -m venv .venv
source .venv/bin/activate
pip install -e ".[dev]"
pytest   # 111 passed, 2 skipped
llm4osc score
llm4osc score --suite paraphrase
\end{lstlisting}

For Track~C (B1--B3): \texttt{pip install -e ".[dev,llm]"}, then \texttt{llm4osc serve} and \texttt{llm4osc score-compare --backends b0,b1,b2,b3}.

\bibliographystyle{plainnat}
\bibliography{llm4osc}

\end{document}